\documentclass{pasj01}
\twocolumn
\Received{$\langle$reception date$\rangle$}
\Accepted{$\langle$acception date$\rangle$}
\Published{$\langle$publication date$\rangle$}

\begin{document}

\title{ Solar horizontal flow evaluation using neural network and numerical simulation with snapshot data}
\author{Hiroyuki MASAKI$^{1, 2}$, Hideyuki HOTTA$^{2, 1}$ Yukio KATSUKAWA$^{3}$ \& Ryohtaroh T. Ishikawa$^{2, 3, 4}$}
\altaffiltext{}{$^{1}$Department of Physics, Graduate School of Science, Chiba University, 1-33 Yayoi-cho, Inage-ku, Chiba 263-8522, Japan \\
$^2$ Institute for Space-Earth Environmental Research, Nagoya University, Furo-cho, Chikusa-ku, Nagoya, Aichi 464-8601, Japan\\
$^3$National Astronomical Observatory of Japan, 2-21-1 Osawa, Mitaka, Tokyo, 181-8588, Japan \\
$^4$National Institute for Fusion Science, 322-6 Oroshi-cho, Toki, Gifu 509-5292, Japan
}
\email{afpa6720@chiba-u.jp}


\maketitle

\begin{abstract}
  We suggest a method that evaluates the horizontal velocity in the solar photosphere with easily observable values using a combination of neural network and radiative magnetohydrodynamics simulations. All three-component velocities of thermal convection on the solar surface have important roles in generating waves in the upper atmosphere. However, the velocity perpendicular to the line of sight (LoS) is difficult to observe. To deal with this problem, the local correlation tracking (LCT) method, which employs the difference between two images,  has been widely used, but LCT has several disadvantages. We develop a method that evaluates the horizontal velocity from a snapshot of the intensity and the LoS velocity with a neural network. We use data from numerical simulations for training the neural network. While two consecutive intensity images are required for LCT, our network needs just one intensity image at only a specific moment for input. From these input array, our network outputs a same-size array of two-component velocity field.
  With only the intensity data, the network achieves a high correlation coefficient between the simulated and evaluated velocities of 0.83.
  In addition, the network performance can be improved when we add LoS velocity for input, enabling achieving a correlation coefficient of 0.90.
  Our method is also applied to observed data.

\end{abstract}

\section{Introduction}
The solar surface is filled with turbulent thermal convection. The energy is continuously generated by nuclear fusion around the centre of the sun. This input energy is transported outward by the radiation in the radiation zone (70\% of solar radius). In the outer 30\% of the solar interior, the energy is transported outward by thermal convection. This layer is called the convection zone (e.g., \cite{2009LRSP....6....2N}). This thermal convection causes a mottled appearance called granulation at the surface. The lifetime, the spatial scale, and the typical velocity of the granulation are several minutes, 1 Mm, and 3--4 $\mathrm{km\ s^{-1}}$, respectively (e.g., \cite{1990ARA&A..28..263S}). Thermal convection in the solar photosphere causes several phenomena in the upper atmosphere and is related to poorly understood solar phenomena, such as coronal heating and magnetic field generation. Thus, it is important to evaluate the thermal convection velocity in the sun. The line of sight (LoS) velocity (i.e., the Doppler velocity) can be measured by the Doppler effect. For example, satellites for solar observation, such as Hinode \citep{2007SoPh..243....3K,2007AAS...210.9405T} and the Solar Dynamics Observatory (SDO:\cite{2012SoPh..275....3P,2012SoPh..275..207S}), have instruments for observing the Doppler shifts at multiple spectral lines.

While we can evaluate the LoS flow velocity using the Doppler effect relatively easily, the flow velocity perpendicular to the LoS is difficult to measure because the motion does not cause any Doppler shift. To deal with this problem, local correlation tracking (LCT:\cite{1988ApJ...333..427N}) is widely used. This method evaluates the horizontal velocity field from the displacements of structures within two successive intensity maps. Because LCT compares close sub-region pairs and finds a large correlation between the two images, this method requires many numerical operations. Moreover, LCT cannot be used with images of arbitrary cadence. In addition, LCT can only evaluate the mean in time and is not good at detecting steady flow in which no apparent motion can be observed.

By contrast, many magnetohydrodynamics (MHD) simulations of the solar photosphere have been improved in the past three decades \citep{1998ApJ...499..914S,2005A&A...429..335V}. The improvements in computer performance and algorithms make it possible to reproduce solar observations in simulations in detail. We can obtain many sets of data such as intensity and three-component velocity in each snapshot using numerical simulations. Numerical simulations have also been used to validate LCT \citep{2013A&A...555A.136V}.

\citet{2017A&A...604A..11A} developed an algorithm that estimates the horizontal velocity field using a combination of numerical simulation and neural network as a substitute for LCT. This algorithm, DeepVel, estimates horizontal velocity at optical depth $\tau$=1, 0.1, and 0.01 from two intensity maps obtained 30 seconds apart. The DeepVel obtains a correlation coefficient of 0.83 at $\tau$=1 between the estimated and simulated velocity. In addition, DeepVelU 
\citep{2020FrASS...7...25T}
, an enhanced version of DeepVel, can use the intensity, the LoS velocity field, and LoS magnetic field as trackers and achieve a correlation coefficient of 0.947. These algorithms also achieved similar values of the correlation coefficient at the other optical depths and can detect vortices more clearly than LCT. The DeepVel and DeepVelU evaluate the horizontal velocity from two images at a specific interval. When the cadence of the new data is different from that used in training, the network needs to be trained again for the new data.
Moreover, \citet{2022A&A...658A.142I} improved the correlation coefficient to 0.95 using a network structure focusing on spatial scales.

In this study, we perform numerical simulations to obtain modeled physical quantities and develop a method that estimates the horizontal velocity field in the solar surface from the intensity and the LoS velocity in one observation snapshot with the neural network using the calculated data. Because the network is constructed only with convolution, the network evaluation is fast for any intensity image size. 
A big advantage of this study compared with the previous research (e.g., \cite{2017A&A...604A..11A}) is that we only require a single snapshot for the evaluation. Thus, we can apply the network to observations with any length of the time cadence. 
We confirm that the network can be applied to observations and compare our result with that of LCT.

\section{neural network training}

\subsection{Numerical simulation}
The data used in this study are calculated by the Radiation and RSST for Deep Dynamics (R2D2:\cite{2019SciA....5.2307H,2020MNRAS.494.2523H,2020MNRAS.498.2925H}) MHD simulation code. The R2D2 solves the following equations.

\begin{eqnarray}
   \frac{\partial\rho}{\partial t}
   &=&
   -\nabla\cdot(\rho \boldsymbol{v})
\\
   \frac{\partial }{\partial t}(\rho \boldsymbol{v})
   &=&
   -\nabla\cdot(\rho \boldsymbol{v}\boldsymbol{v})
   -\nabla p
   +\rho \boldsymbol{g}
   +\frac{1}{4\pi}
   \left(\nabla\times\boldsymbol{B}\right)\times \boldsymbol{B}
\\
   \frac{\partial \boldsymbol{B}}{\partial t}
   &=&
   \nabla
   \times\left(\boldsymbol{v}\times\boldsymbol{B}\right)
\\
   \rho T \frac{\partial s}{\partial t}
   &=&
   \rho T \left(\boldsymbol{v}\cdot\nabla\right)s + Q
\\
   p &=& p\left(\rho,:s\right)
\end{eqnarray}
Here, $\rho$, $\boldsymbol{v}$, $p$, $T$, $\boldsymbol{g}$, $\boldsymbol{B}$, $s$, and $Q$ are the density, velocity, pressure, temperature, gravitational acceleration, magnetic field, entropy, and radiative heating, respectively. The R2D2 solves the equations with a fourth-order spatial derivative and four-step Runge--Kutta method for time integration. Pressure $p$ is obtained from the entropy and the density table prepared by the OPAL equation of state considering partial ionisation \citep{1996ApJ...456..902R}. Radiative heating $Q$ is calculated by solving the radiation transfer equation using the grey approximation and the short characteristic in 24 directions.

The simulation box size is 6.144 Mm $\times$ 6.144 Mm in the horizontal direction and 3.072 Mm in the vertical direction. The number of grid cells is 128 in each direction. Thus, the horizontal and vertical grid spacings are 48 km and 24 km, respectively. Considering the typical lifetime of the granulation (several minutes), we set the output cadence to five minutes. While we can obtain more data with a shorter cadence, the convection structure does not change significantly with the shorter time interval. We choose the output cadence to compromise the data amount and the neural network training efficiency. We initiate calculations with different initial vertical magnetic fields, 1, 20, and 30 G, to ensure data generality. We obtain about 30,000 snapshots of data. For the training, we use the data at $\tau$=1 surface defined with the Rosseland mean opacity.

\subsection{Network structure}
In this study, we mainly train two neural networks named Networks I and IV. The radiative intensity is used as an input for both networks, and the vertical velocity at the $\tau=1$ surface is used as additional input for Network IV. These networks have almost the same structure. We emphasize that we use a huge amount of data for the training, but we only require a single snapshot of the data for the practical evaluation. The output is two components of the horizontal velocity with the same number of grid points as the original input data. The network has the encoder--decoder structure only with convolution. This structure is often used in image recognition because it can be applied to any image size by a learning filter. The encoder--decoder structure is divided into encoder and decoder. The encoder extracts the feature of the input image and compresses the image, and the decoder converts the compressed image to an output image. In this study, the network is based on U-net \citep{2015arXiv150504597R} with some skip connection that delivers information from the encoder to the decoder. The common encoder--decoder tends to lose positional information on input. U-net solves this problem using the skip connection. In addition, we deepen the networks by placing ResidualNet between the encoder and decoder. ResidualNet is a structure that repeats the skip connection at a small interval. 

\begin{figure}
  \centering
  \includegraphics[width=\linewidth, clip]{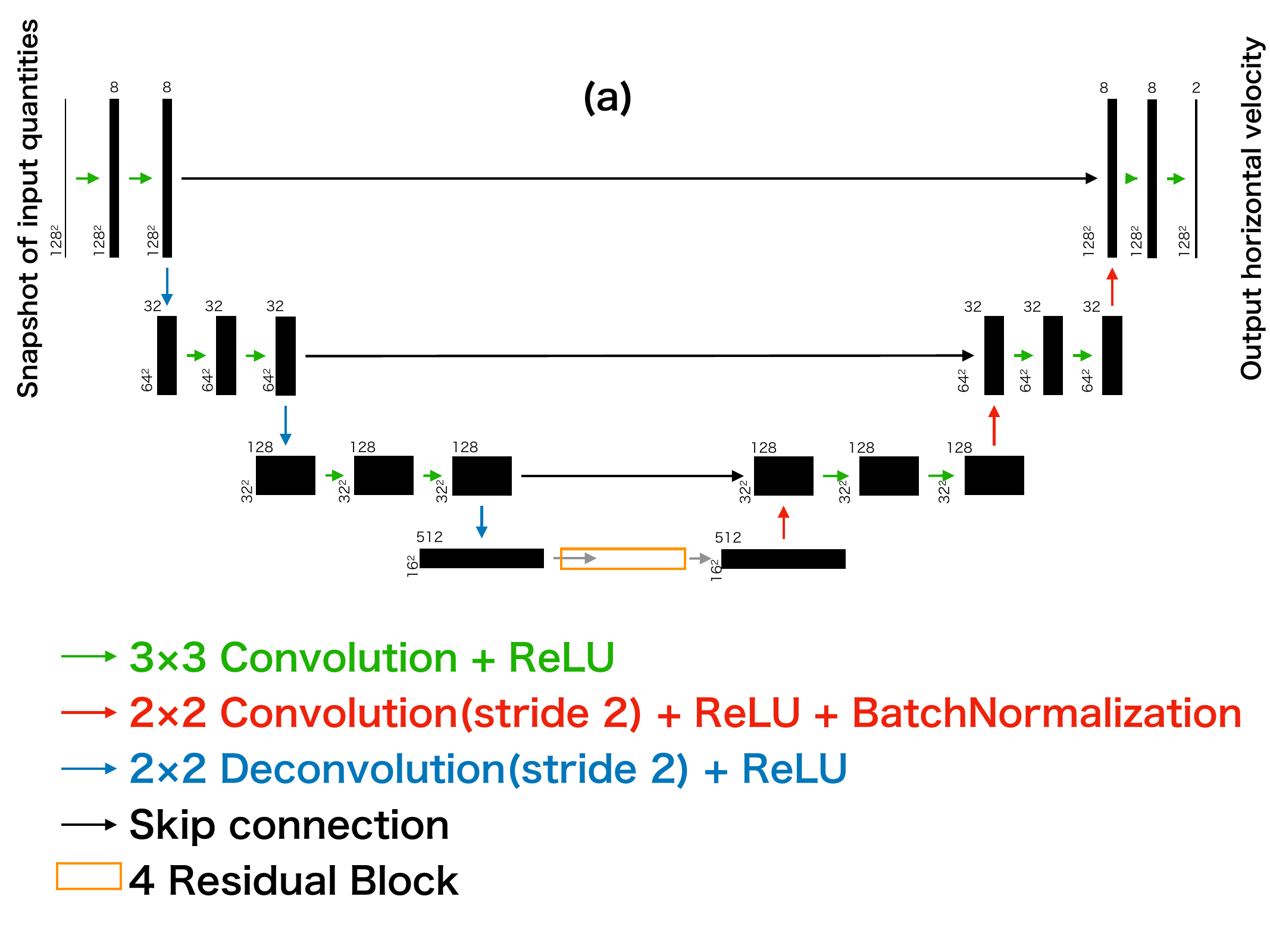}
  \includegraphics[width=\linewidth, clip]{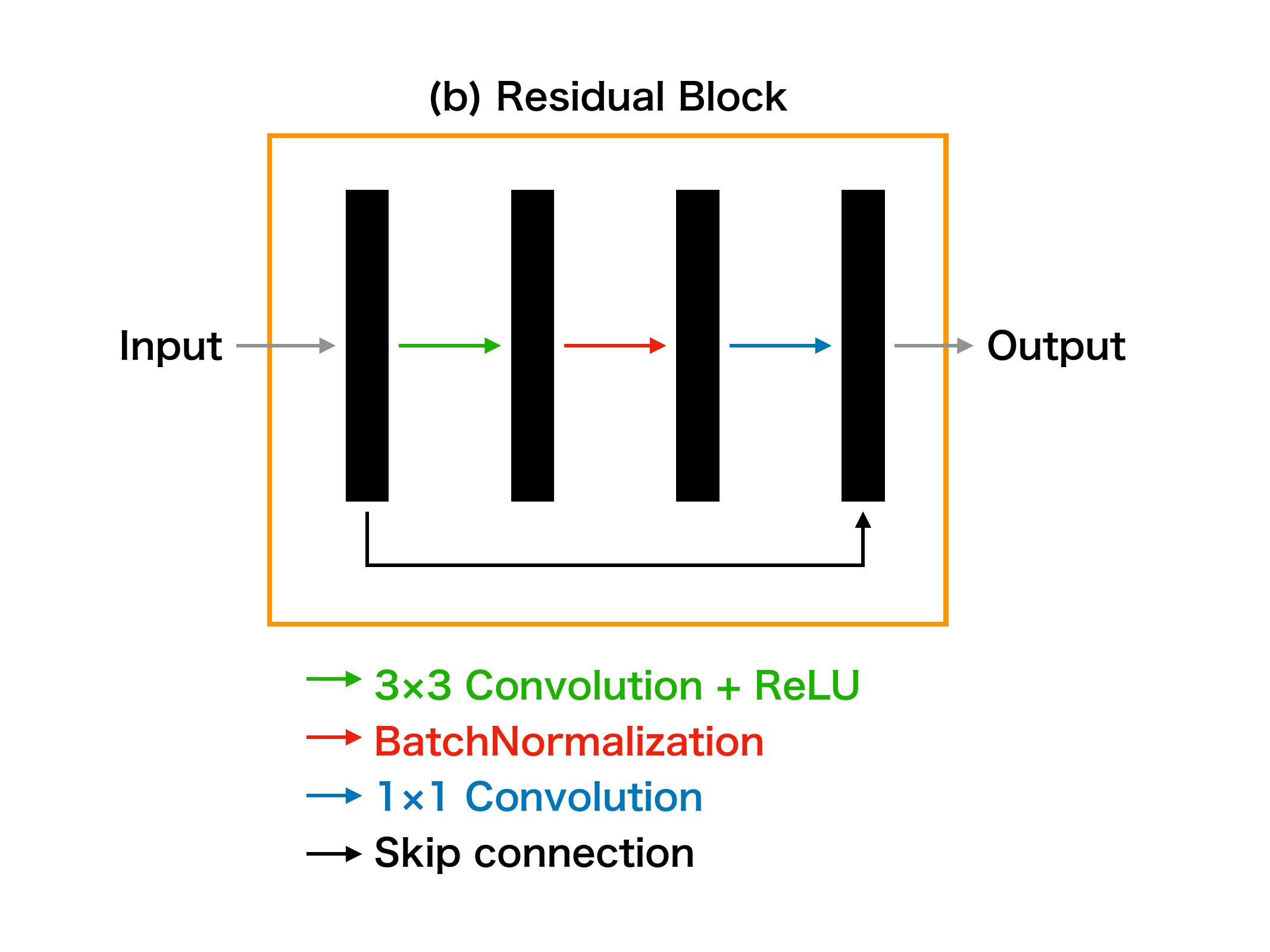}
  \caption{The network architecture is shown. The top panel shows the whole network. The bottom panel shows the residual network in the orange box in the top panel. The black squares show the shape of the data, and the numbers on the left and the top show the size and number of the images, respectively. The process shown by the coloured arrows change the shape of the data. The input and output of the residual blocks are the same.}
  \label{NA}
\end{figure}

The network structure is shown in Fig. \ref{NA}. 
The network can be applied to any size of two-dimensional input intensity and vertical velocity image. We note that some layers divide the width of an image in half, and bit error may increase with an odd number of pixels. The kernel size of convolution is $3\times3$, and we use rectified linear unit (ReLU) for the activation function. When image size is reduced by downsampling, we quadruple the number of filters so that the amount of information does not decrease. The U-net was developed for biological-image segmentation. In the segmentation problem, max pooling is generally used to extract features of the image. All information of the input image in this study is related to output images (horizontal velocities). Thus, we choose the convolution with the stride size of two and a kernel size of $2\times2$ for downsampling. We use the same size, stride, and number of filters as the convolution process in the deconvolution process, i.e., the reversed procedure of convolution is adopted for the deconvolution. We note that in the final step, we output two images of velocity (two components of velocity). In our network, we obtain a set of velocity fields in a square area 4 Mm on a side.
\subsection{Training setting}
We use the intensity and vertical velocity maps of $128\times128$ pixels obtained from the numerical simulations as the input of the network. The intensity is normalised by the temporal and spatial average intensity of the sun. For outputs, we use the two-component velocity field of $128\times128$ pixels. The unit for velocities is $\mathrm{km\ s^{-1}}$.
We prepare about 30,000 datasets for training. We also rotate the image by 90, 180, and 270 degrees to increase the amount of data. We note that the DeepVel adopted the same way, i.e., the rotated image, for increasing the data amount \citep{2017A&A...604A..11A}. Finally, we use about 120,000 datasets for training. In addition to these data, we prepare about 1,000 datasets for network validation. Mini-batches comprising 32 datasets are randomly selected from training data. We note that the mini-batch is a set of datasets, with which the network performance tends to increase \citep{2018arXiv180308494W}. The network trains in 128 epochs, and the network with the highest correlation coefficient between the evaluation of the network and validation data is adopted. The network is optimised to minimise the mean square error using the Adam optimiser \citep{2014arXiv1412.6980K}.

We use TensorFlow and its wrapper, Keras, for implementing the network and an Nvidia GeForce RTX 2080 Ti GPU for training.

\section{Result}
\subsection{Validation of image}

\begin{figure*}
  \centering
  \includegraphics[width=\linewidth, clip]{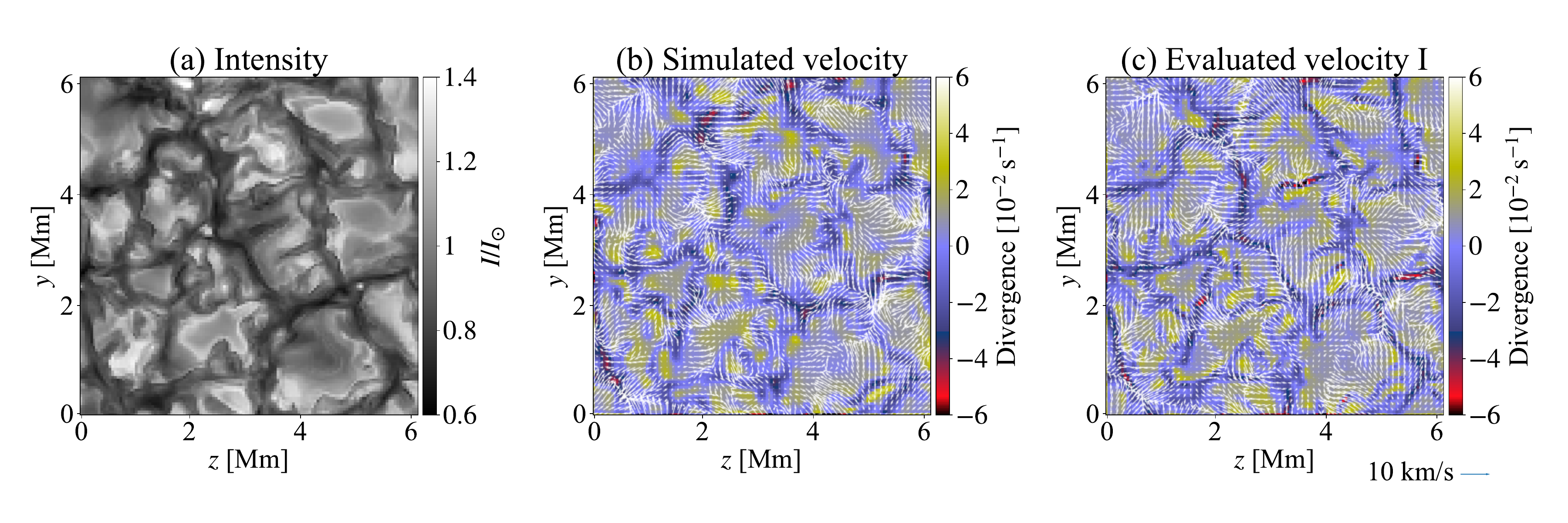}
  \includegraphics[width=\linewidth, clip]{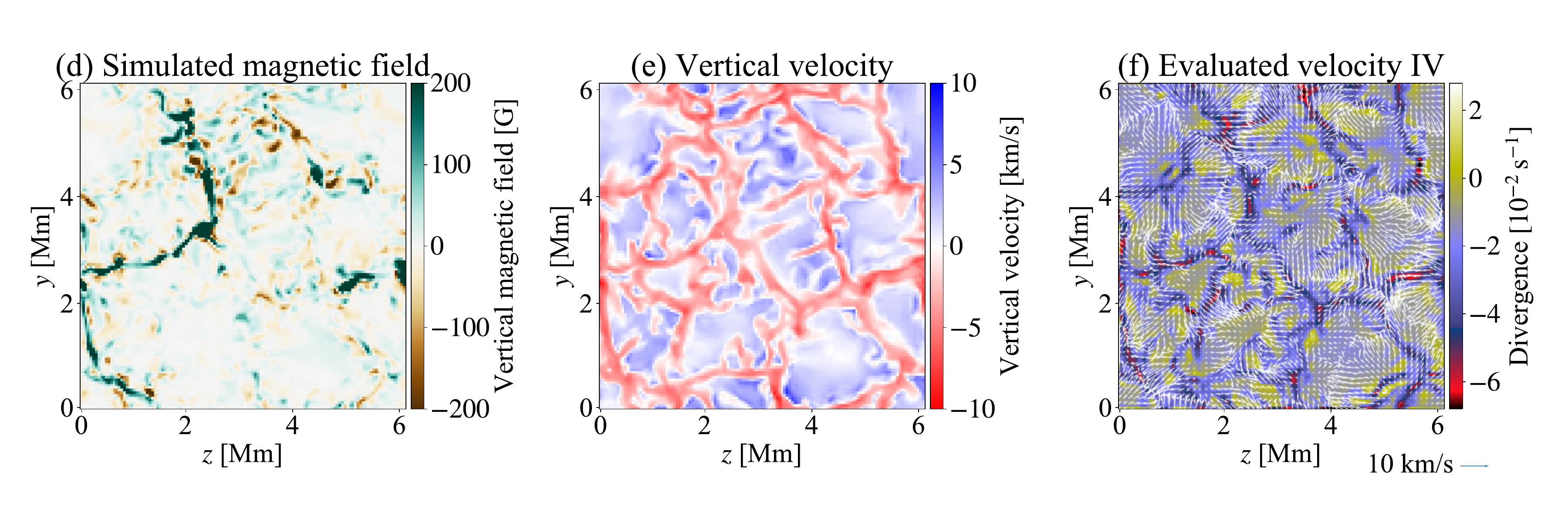}
  \caption{Examples of the evaluation: (a) intensity, (b) simulated horizontal velocity, and (c) estimated horizontal velocity by Network I, and (d) vertical magnetic field, (e) vertical velocity, and (f) estimated horizontal velocity by Network IV are shown. The white arrows show the velocity, and an example of a length of 10 $\rm{km\: s^{-1}}$ is shown by the blue arrow in the lower right corner. Note that the input value for the magnetic field is preprocessed and thus differs from (d).}
  \label{arrow}
\end{figure*}

Fig. \ref{arrow} shows the estimated horizontal velocity by Networks I and IV. Panels a, b, c, d, e, and f show the intensity, simulated horizontal velocity, and estimated horizontal velocity by Network I, and the vertical magnetic field, vertical velocity, and estimated horizontal velocity by Network IV, respectively. The white arrows show the horizontal velocity, and the background colour map shows the horizontal divergence of the flow. We use the magnetic field in \S \ref{sec:dependence_on_input}. The results show that the structure estimated by the network roughly reproduces the original velocity field. The network can also detect the vortex at $(y, z)=(2\ \mathrm{Mm}, 5\ \mathrm{Mm})$.

We calculate the correlation coefficient defined as:
\begin{equation}  
  \frac{\sum(v_{\rm sim}-\bar v_{\rm sim})(v_{\rm eva}-\bar v_{\rm eva})}
  {\sqrt{\sum(v_{\rm sim}-\bar v_{\rm sim})^2}\sqrt{\sum(v_{\rm eva}-\bar v_{\rm eva})^2}}
\end{equation}
where $v_\mathrm{sim}$ and $v_\mathrm{eva}$ are the simulated and estimated horizontal velocity for the validation dataset, respectively. $\sum$ is the sum of all pixels and all validation data, and the overbar is the mean of the two-dimensional data. The correlation coefficients for Networks I and IV are 0.83 and 0.90, respectively. The mean absolute error values
\begin{equation}
  \overline{|\vec v_{\rm sim}- \vec v_{\rm eva}|}
\end{equation}
are 0.92 $\rm{km\: s^{-1}}$ and 0.72 $\rm{km\: s^{-1}}$, the R2score values
\begin{equation}
1 - 
\frac
{\sum (v_{\rm sim}- v_{\rm eva})}
{\sum (v_{\rm sim}-\bar v_{\rm sim})}
\end{equation}
are 0.71 and 0.82,
and the mean angular differences
are $28.6^\circ$ and $22.6^\circ$
for Networks I and IV, respectively.
The mean angular difference $\theta$ describes the angle between the flow vectors of the evaluated and simulated flows.
\begin{equation}
  \theta = \arccos
  \left(
  \frac{\boldsymbol{v}_\mathrm{eva}\cdot \boldsymbol{v}_\mathrm{sim}}
  {|\boldsymbol{v}_\mathrm{eva}||\boldsymbol{v}_\mathrm{sim}|}
  \right)
\end{equation}

\begin{figure}
  \centering
  \includegraphics[width=\linewidth, clip]{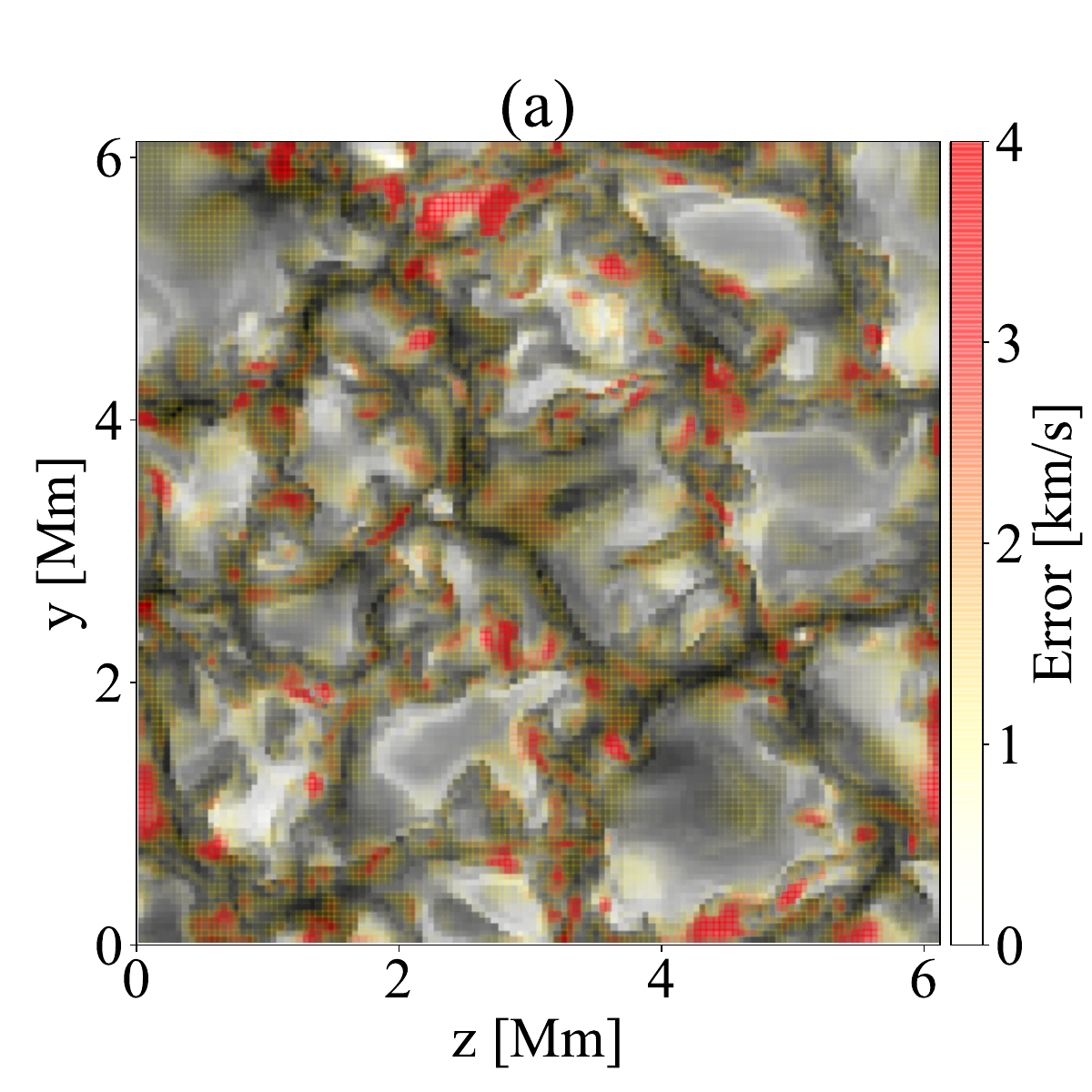}
  \includegraphics[width=\linewidth, clip]{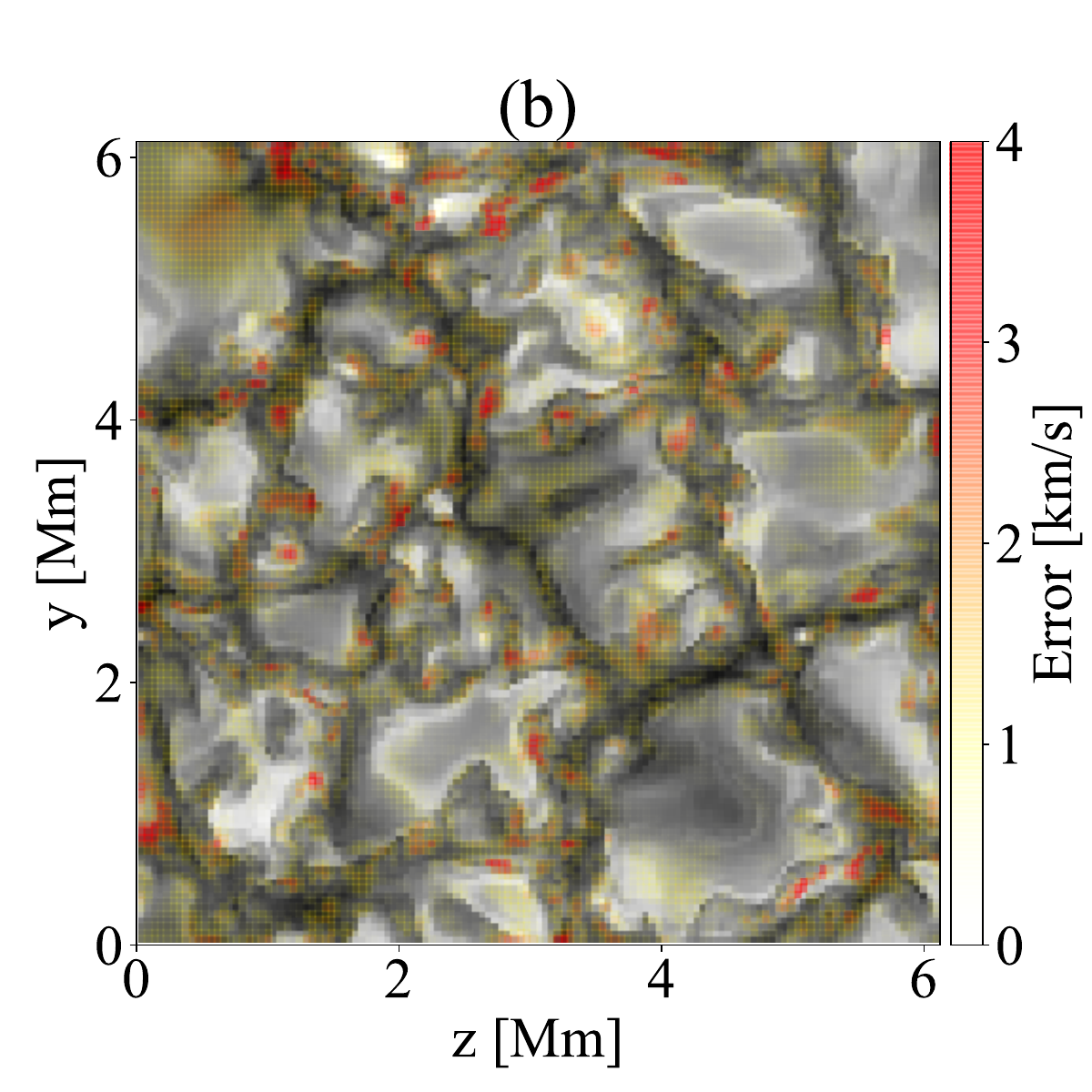}
  \caption{(a) and (b) The differences between the simulated and estimated velocities from Networks I (intensity) and IV (intensity and vertical velocity), respectively. The background grey map shows the input intensity.
  }
  \label{deff}
\end{figure}
Fig. \ref{deff} shows the difference in the absolute value between the simulated and evaluated values. Panels a and b show the results from Networks I and IV, respectively. One may notice that the difference increases around the granular boundary. This is because the structure of the granule boundary is complex, and the sign of velocity changes in this area. This small-scale turbulence does not obey the overall coherent pattern of thermal convection, i.e., the broad areas of diverging flows and narrow lanes of converging flows.

\begin{figure}
  \centering
  \includegraphics[width=\linewidth, clip]{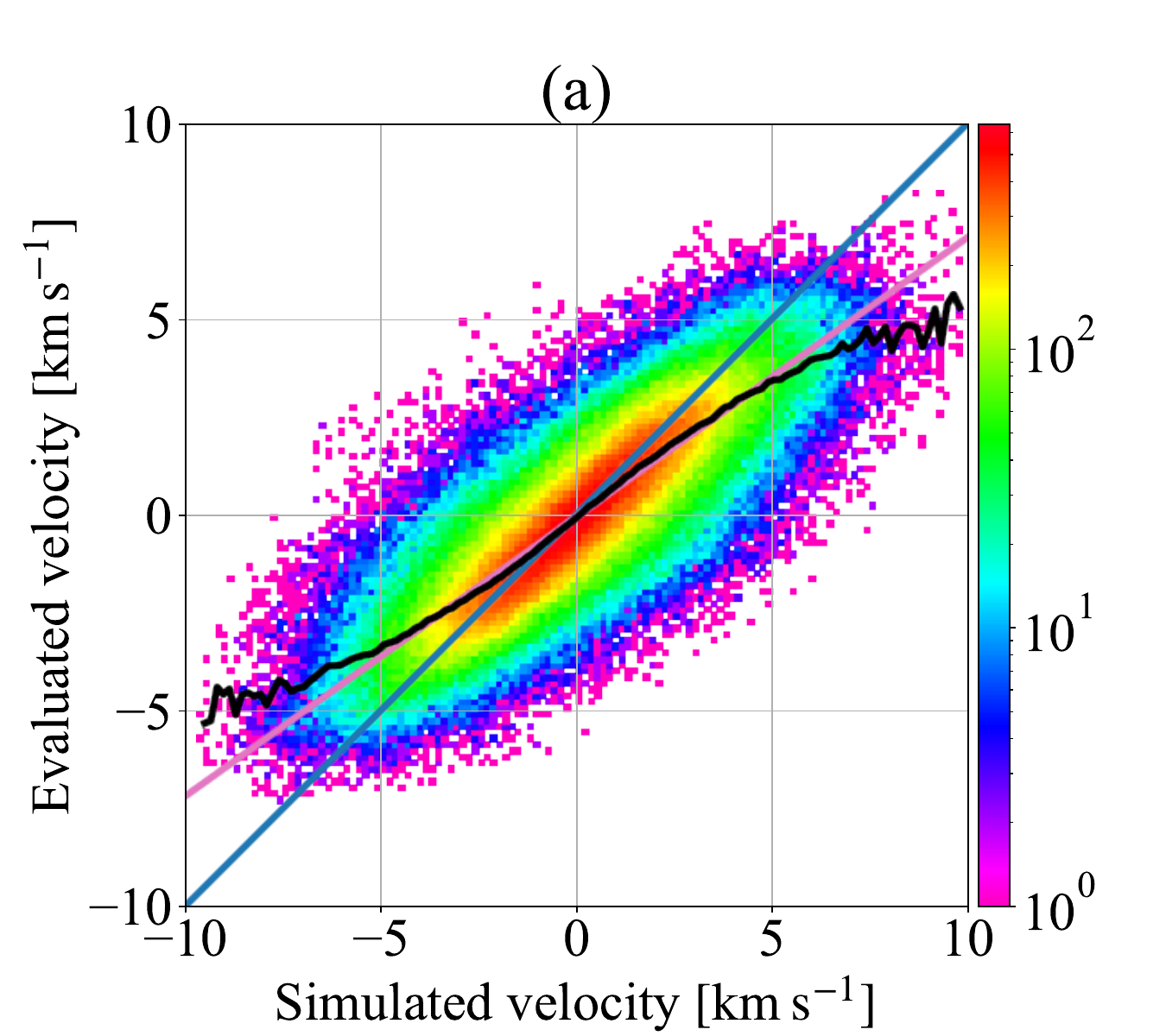}
  \includegraphics[width=\linewidth, clip]{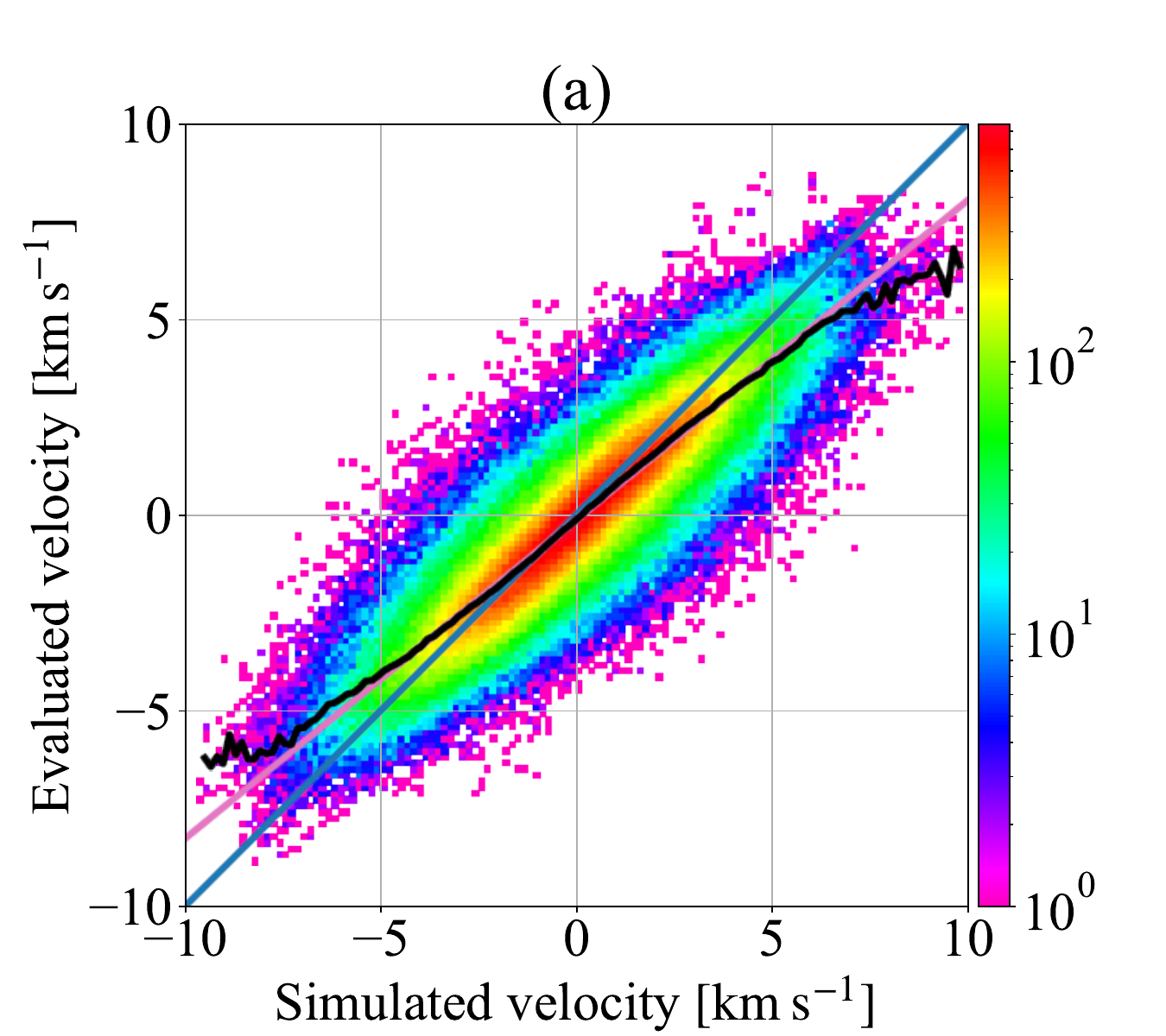}
  \caption{Two-dimensional histogram between the simulated and estimated (evaluated) velocities is shown. (a) The result from Network I and (b) the result from Network IV. The colour map shows the number of pixels that have the corresponding value. The light blue line shows where the two values match exactly $v_\mathrm{sim}=v_\mathrm{eva}$. The pink line shows the result of the linear fitting of the data, where $v_{\rm eva} =0.710v_{\rm sim}-0.076 $ in (a) and $v_{\rm eva} =0.813v_{\rm sim}-0.098 $ in (b). The black line shows the average of the estimated velocities for each simulation velocity.} 
  \label{hist}
\end{figure}

Fig. \ref{hist}
shows two-dimensional histograms of the simulated and estimated velocities. Panels a and b show the results from Networks I and IV, respectively. While most data points are located on the line $y=x$, the network tends to underestimate the velocity with respect to high simulated velocities and the fitted slope is smaller than unity. This result indicates that our evaluation tends to show lower velocity than the simulated velocity.
\subsection{Dependence of networks on the amount of data}
\begin{figure}
  \centering
  \includegraphics[width=\linewidth, clip]{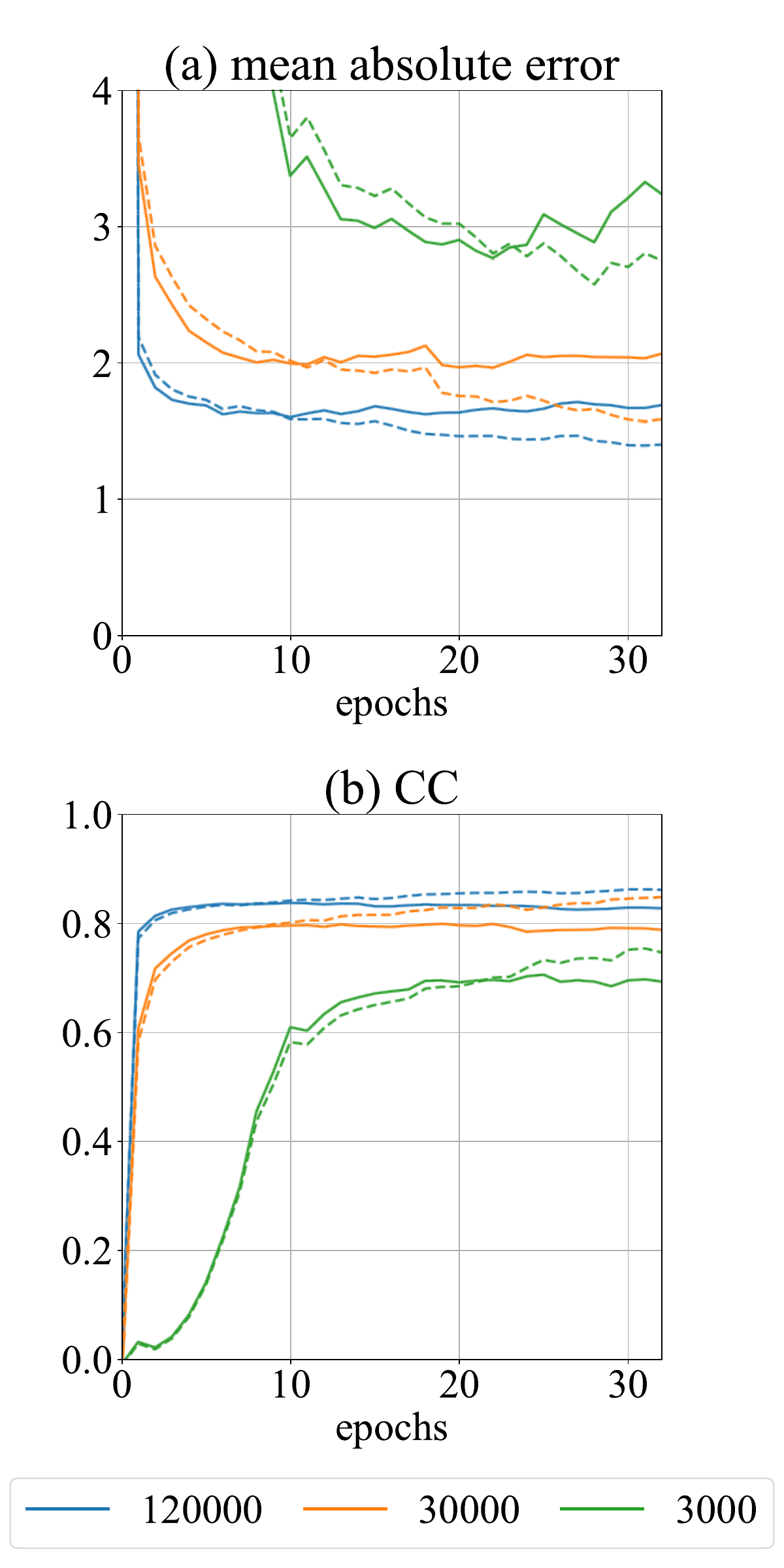}
  \caption{The learning curves with different numbers of datasets are shown. Panels (a) and (b) show the mean absolute error and the correlation coefficient, respectively. The solid line shows the evaluation value for the test data, and the dashed line shows the evaluation value for the training data. The horizontal axis shows the epochs. Note the axis is not the iteration of the network. The gap between the training and test data curves reflects the network's overtraining.
  Note that the trend does not change up to 128 epochs, so this figure only shows 30 epochs.
  }
  \label{learn}
\end{figure}

Overtraining occurs when the training data lack generality and can be suppressed by increasing the amount of data. To test the effect of the amount of data on the network, we trained Network I with different numbers of datasets. The training results with dataset numbers of 3,000, 30,000, and 120,000 are shown in Table \ref{table}.   
\begin{table*}[t]
  \centering
  \tbl{The results with different numbers of datasets}{%
   \begin{tabular}{|l|c|c|c|} 
    \hline
    Number of datasets & 3,000 & 30,000 & 120,000  \\ 
    \hline
    Correlation coefficient & 0.70 & 0.80 & 0.83  \\
    R2score & 0.42 & 0.63 & 0.69  \\
    Mean square error $\rm [(km\,s^{-1})^2]$ & 1.76 & 1.41 & 1.29 \\
    Mean absolute error $\rm [km\,s^{-1}]$& 1.36 & 1.06 & 0.96  \\ 
    Mean angle [degree] & $45.0^\circ$ & $33.7^\circ$ & $29.7^\circ$ \\
    \hline
   \end{tabular}}
   \begin{tabnote}
    \hangindent6pt\noindent
    \hbox to6pt{\footnotemark[$*$]\hss}\unskip%
    The results of validation functions for Network I.
  \end{tabnote}
  \label{table}
\end{table*}
We note that the correlation coefficient and R2score should be unity for the perfect evaluation. The other evaluation parameters should be zero for the perfect evaluation.  Fig. \ref{learn}
also shows the learning curves of these data. These results show that increasing the amount of data improves the network performance and suppresses overtraining. The solid and dashed lines show the learning curve for training and validation data, respectively. The discrepancy between these lines indicates overtraining. The estimation ability does not change much between 30,000 and 120,000 datasets. This indicates that increasing the amount of data does not drastically improve network performance. 

\subsection{Dependence of networks on input}\label{sec:dependence_on_input}
\begin{figure}
  \centering
  \includegraphics[width=\linewidth, clip]{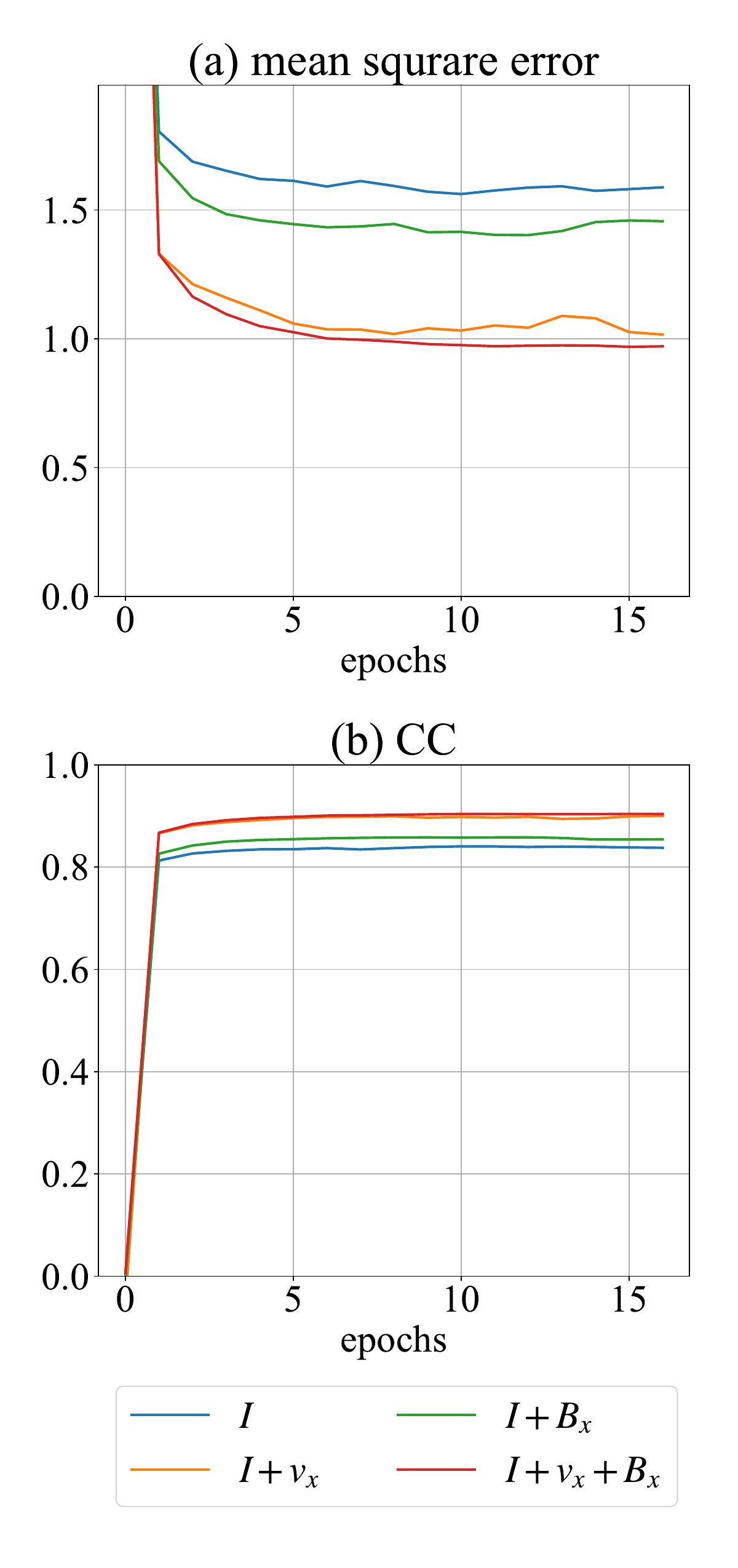}
  \caption{The learning curves of the networks that require different inputs is shown. (a) The mean absolute error and (b) correlation coefficient.}
  \label{input}
\end{figure}

In the previous sections, we compare Networks I (intensity) and IV (intensity and vertical velocity). In this section, we perform additional investigations for the input. We use a vertical magnetic field that is observable by the Zeeman effect as input data. We train four networks using different inputs: (I) intensity, (IV) intensity and vertical velocity, (IB) intensity and vertical magnetic field, and (IVB) intensity, vertical velocity, and magnetic field. Because intensity is the easiest to observe, we do not consider the network without intensity. Fig. \ref{arrow} shows the example of physical quantities to input in panels a, d, and e. The distribution of the magnetic field has a large kurtosis, in which most of the values are close to zero. This type of distribution is difficult to use for evaluation. Thus, we process the magnetic field as:
\begin{equation}
  B_x' = \frac{B_x}{|B_x|}\log\left(1+\frac{B_x}{B_{\rm cr}}\right),
\end{equation}
where $B$ and $B'$ are the original magnetic field and the processed quantity, respectively. $B_{\rm cr}$ is a free parameter, and we choose $B_{\rm cr}=1\:\rm G$. Table \ref{inputT} and Fig. \ref{input} 
\begin{table*}[t]
  \centering
  \tbl{The results with different inputs}{%
  \begin{tabular}{|l|c|c|c|c|} 
    \hline
    Network                            & I            & IV           & IB           & IVB     \\
    Input data                         & $I$          & $I,v_x$      & $I,B_x$      & $I, v_x, B_x$\\ 
    \hline
    Correlation coefficient            & 0.84         & 0.90         & 0.86         & 0.90\\
    R2score                            & 0.71         & 0.81         & 0.74         & 0.81\\
    Mean square error $\rm [(km\,s^{-1})^2]$ & 1.25         & 1.01         & 1.18         & 0.98\\
    Mean absolute error $\rm [km\,s^{-1}]$   & 0.92         & 0.74         & 0.88         & 0.72\\ 
    Mean angle [degree]                 & $28.6^\circ$ & $22.6^\circ$ & $26.9^\circ$ & $21.9^\circ$\\
    \hline
  \end{tabular}}
  \begin{tabnote}
    \hangindent6pt\noindent
    \hbox to6pt{\footnotemark[$*$]\hss}\unskip%
    The results of validation functions with different inputs. $x$ indicates the vertical direction.
  \end{tabnote}
  \label{inputT}
\end{table*}
show the results of the validation functions of different networks and their learning curves, respectively. The network performance is improved by adding vertical velocity. This is because the velocity field is less diffuse than intensity and has more detailed information. By contrast, the role of the magnetic field is much less important. Especially, the difference between Networks (IV) and (IVB) is insignificant. The effect of the vertical magnetic field on the velocity field is indirect.

\subsection{Comparison with previous studies}
We compare Network I with the DeepVel \citep{2017A&A...604A..11A} and DeepVelU \citep{2020FrASS...7...25T}, which obtain the solar velocity field using a method similar to that used in this study. The DeepVel and DeepVelU estimate the horizontal velocity with two consecutive intensity images, LoS velocity or LoS magnetic field. This means that these methods are an alternative to LCT. This is different from the network in this study using one snapshot.


The training data for the DeepVel are 30,000 images of $50\times50$ pixels and for the DeepVelU are 2,000 images of $48\times48$ pixels.
Our data are 120,000 images of $128\times128$ pixels. The amount of data used in this study is 20 times larger than that used in the DeepVel. The correlation coefficient between the velocity estimated by the DeepVel and the simulated one is 0.83. Because the DeepVel requires two input images, the velocity can be estimated from the temporal difference. By contrast, our network can estimate it from one snapshot and achieves a similar performance to DeepVel without using temporal evolution information.

\section{Application to Observation}
\begin{figure}
  \centering
  \includegraphics[width=\linewidth, clip]{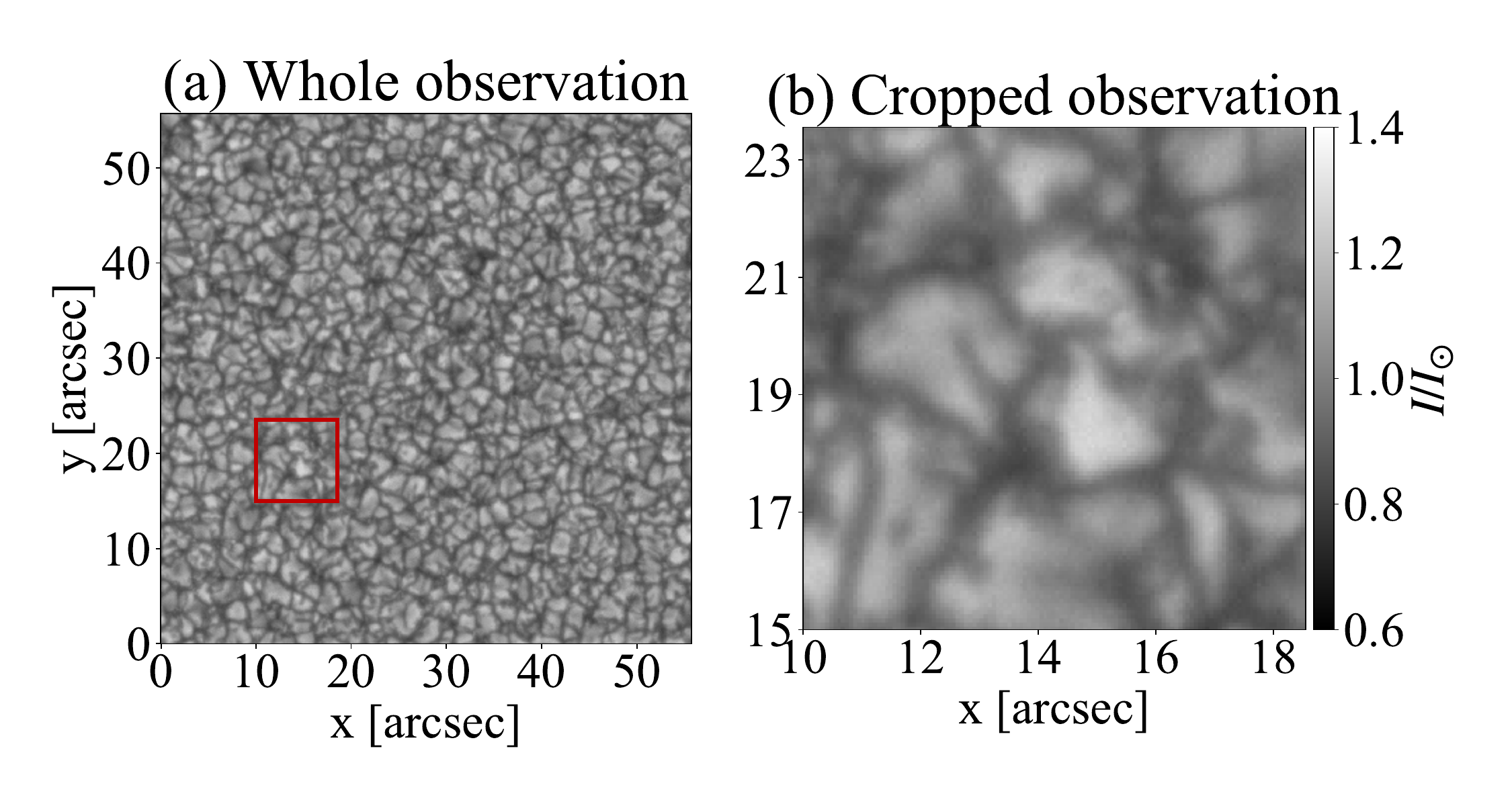}
  \caption{A sample of the Hinode observation images used in this study.
  (a) The entire area of observed data.
  (b) The area enclosed by the red square in (a), with the resolution adjusted using linear completion.
  }
  \label{observation}
\end{figure}

In this section, we apply our network to observed data. The data were taken through a green continuum filter centred at 555 nm by the Hinode Solar Optical Telescope (SOT). We here use a snapshot taken at 11:46:34 on Dec. 29, 2007, with an exposure time of 0.077 s.
Fig. \ref{observation}
shows the overall view of the observed data.
We perform linear interpolation on the 39 km $\times$ 39 km observed data to align the resolution of the 48 km $\times$ 48 km training data. The image with the aligned resolution is shown in Fig. \ref{observation}b, and the area of panel b is indicated by the red square in panel a. The data are normalized by the average radiation intensity of the entire observation image.
The original observation image (a) has a pixel scale of 0.054". We crop out a FOV of about 161 pixels.

\begin{figure}
  \centering
  \includegraphics[width=\linewidth, clip]{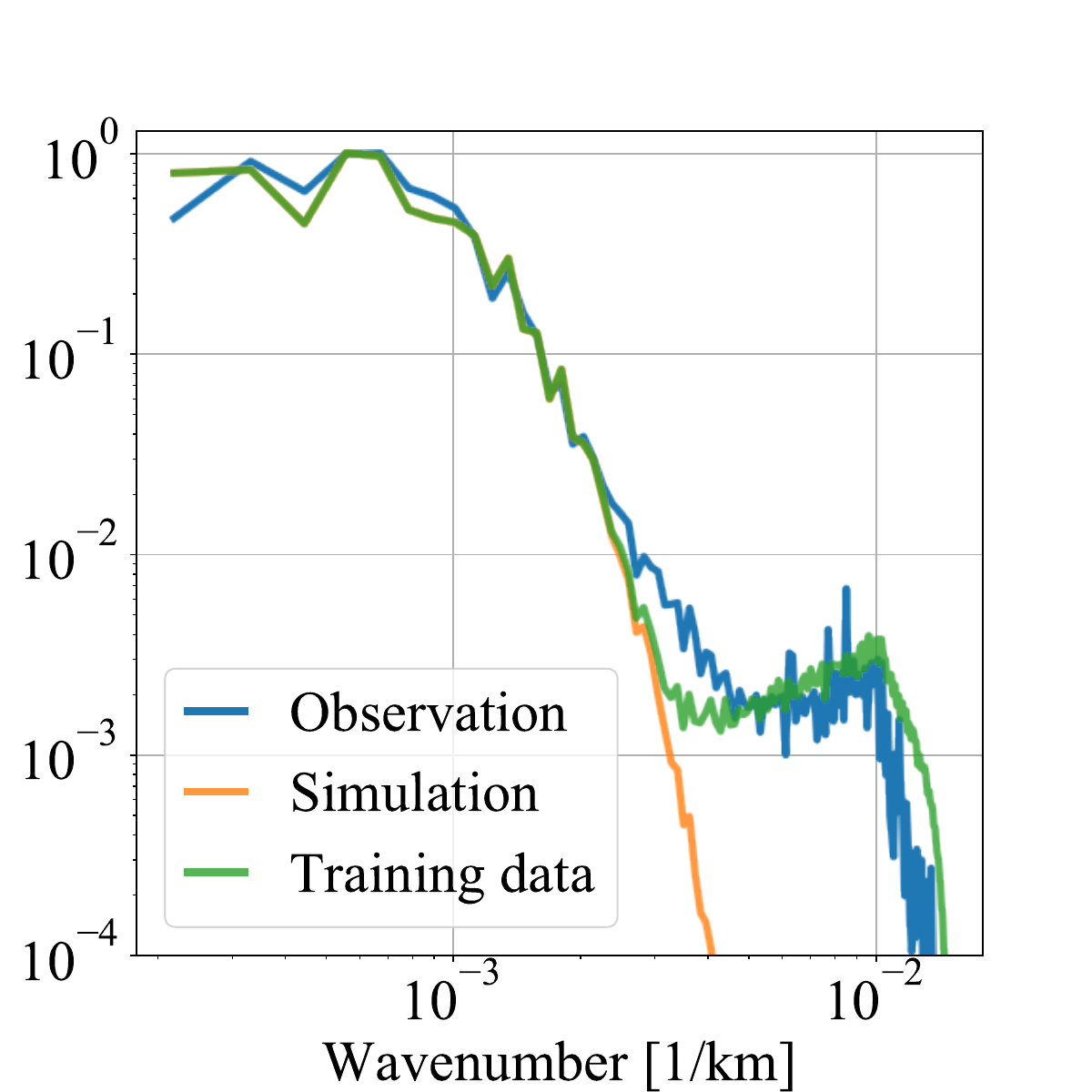}
  \caption{The radiation intensity spectra are shown.
  The blue, orange, and green lines show observation, simulation, and training data, respectively.
  The horizontal axis shows the wavenumber, and the vertical axis shows the intensity normalized by the respective maximum value. }
  \label{wavenumber}
\end{figure}

To apply the network to the observations, we use Hinode's point spread function (PSF) to the intensity of the training images described above. We apply the PSF at 555 nm green continuums described in \citet{2009A&A...501L..19M}. The network trained with the data with the PSF is named Network IP. 
This is not enough to prevent the network from overtraining.
Fig. \ref{wavenumber}
shows the wavenumber distribution of the observed and simulated intensity. While the wavenumber distributions at large scale are consistent, the observation intensity has small-scale structures that are not present in the simulation data. This difference is assumed to be noise in obtaining the horizontal velocities, 
so we add random noise to the data to ignore the small-scale structure of the observed data.
We add random noise in the Fourier space. The mean of the noise is zero and the standard deviation normalized with the maximum intensity is $1.6\times10^{-3}$.
We remove the information of small scales that does not match the observed and simulated data for the training by adding the noise and letting the network apply it. Here we name the network trained with the data with the PSF and noise as Network IPN. Because we intend to evaluate the velocity before using the PSF, we do not apply the PSF for the horizontal velocities, which is the output. All other settings are unchanged from Networks I and IV.

\begin{figure*}
  \centering
  \includegraphics[width=\linewidth, clip]{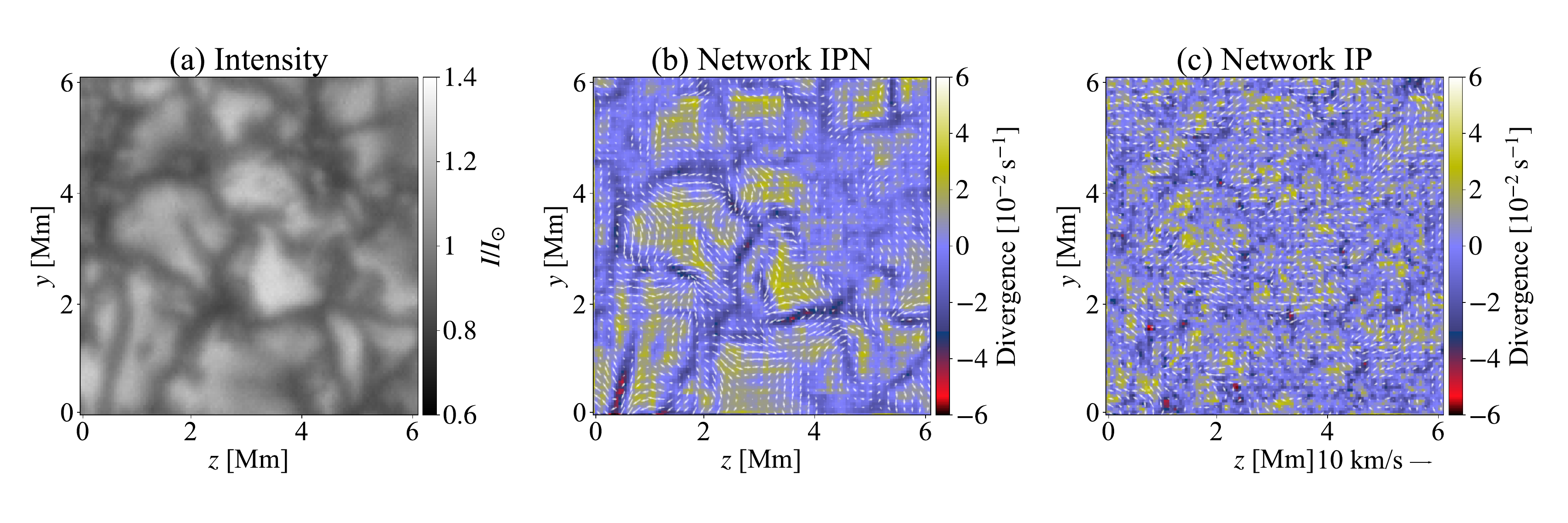}
  \caption{(a) The input intensity observed in the Hinode, (b) the estimation by the network trained with noise, and (c) the estimation by the network trained without noise.
  The result in panel c does not adapt to the Hinode image due to overtraining.}
  \label{over_obs}
\end{figure*}

Fig. \ref{over_obs} shows the results of our application to the observed data. Panel a shows the observed intensity. Panels b and c show the network evaluation of the horizontal velocity by Networks IPN and IP, respectively. Network IP seems to fail the evaluation (panel c) because the velocity structure is not a typical granule network structure, while Network IPN shows reasonable evaluation (panel b). 
This result shows that adding noise is an important factor to apply our network to the real observed data.
\begin{figure*}
  \centering
  \includegraphics[width=\linewidth, clip]{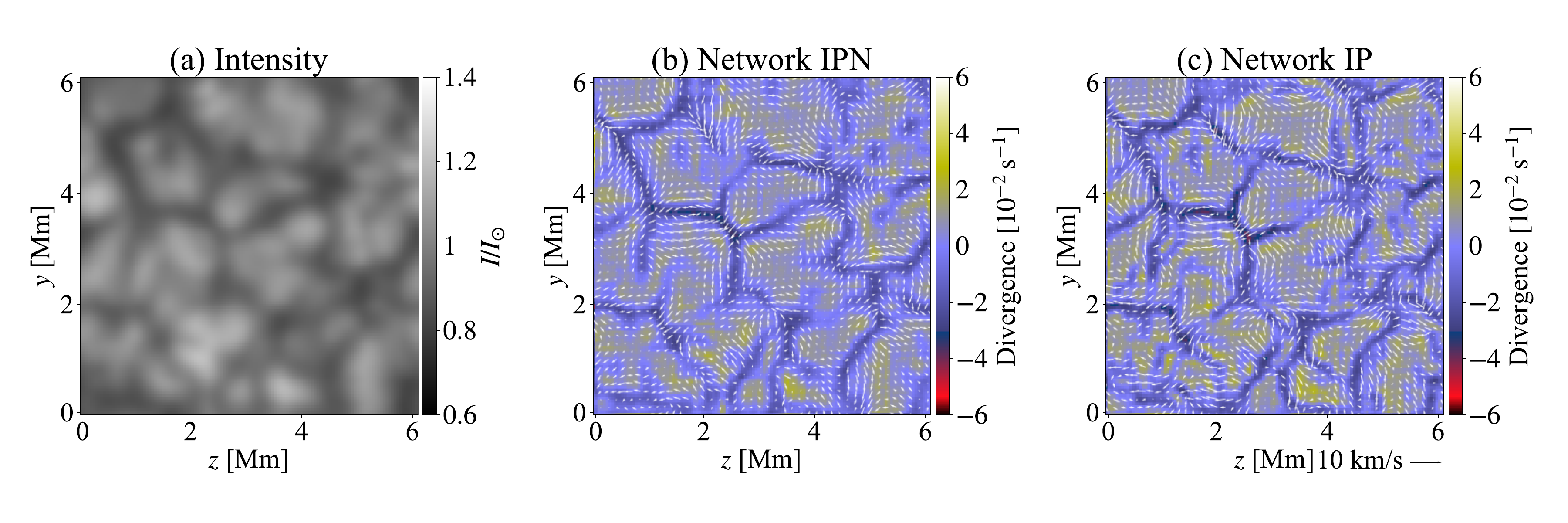}
  \caption{(a) The input intensity in the simulation, (b) the estimation by the network trained with noise, and (c) the estimation by the network trained without noise.}
  \label{over_sim}
\end{figure*} 
We also apply Networks IPN and IP to the simulation data. Figs \ref{over_sim}b and \ref{over_sim}c show the result with Networks IPN and IP, respectively. For the simulation data, the difference between Networks IPN and IP is not significant.
In the early stage of training, the networks can fit the large scale and gradually shift to smaller scales. It becomes difficult for Network IP to estimate velocity from the observed data due to the noise. In Network IPN, the small-scale structure vanishes due to random noise, and thus it can estimate the observations.

\begin{figure}
  \centering
  \includegraphics[width=\linewidth, clip]{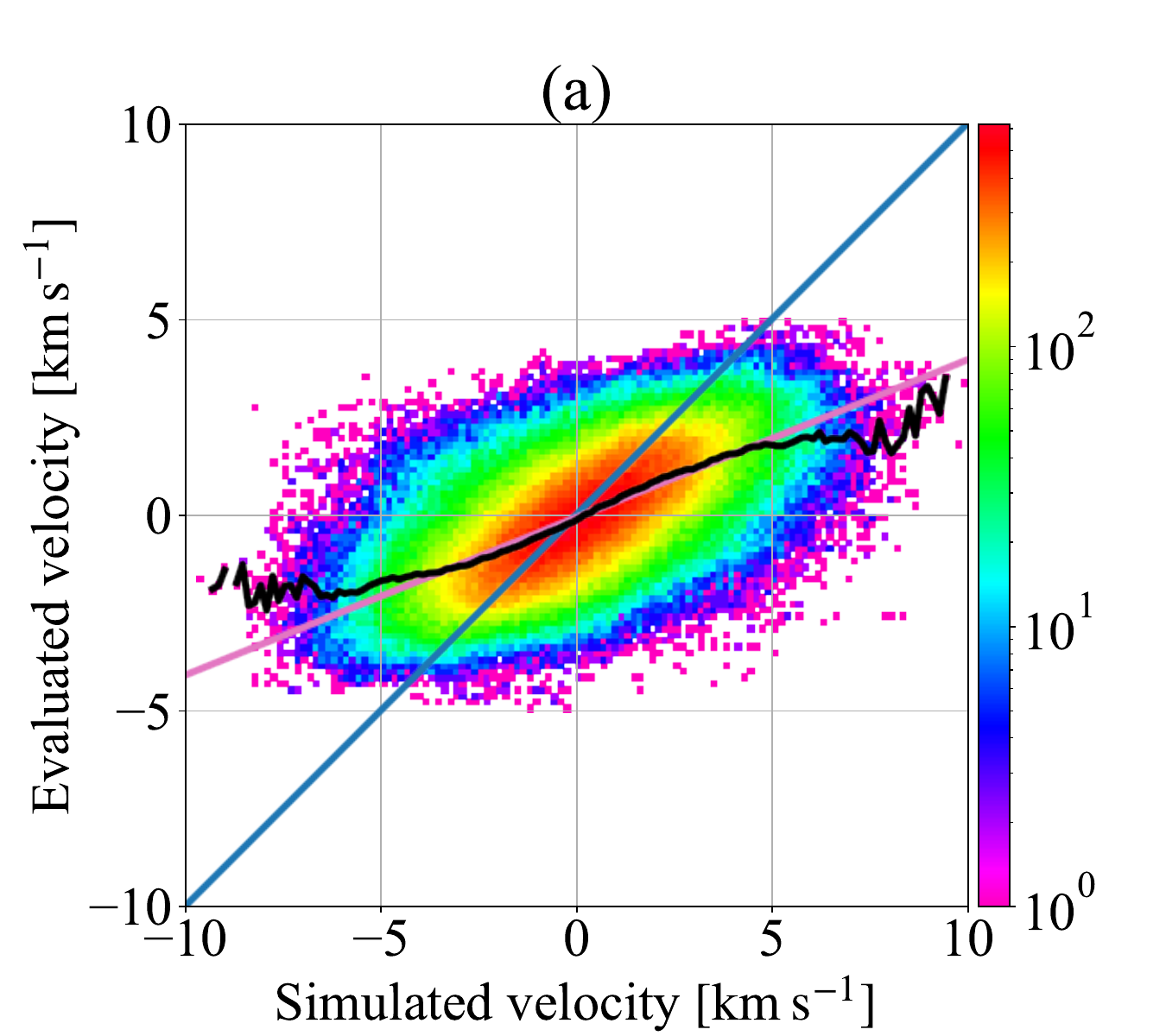}
  \caption{A two-dimensional histogram of the velocity field estimated by Network IPN and the simulated velocity field is shown. The colour map shows the number of pixels that have the corresponding value. The light blue line shows where the two values match exactly $v_\mathrm{sim}=v_\mathrm{eva}$. The pink line shows the result of the linear fitting of the data, where $v_{\rm eva} =0.403v_{\rm sim}-0.067 $. The black line shows the average of the estimated velocities for each simulation velocity.}
  \label{hist_obs}
\end{figure}

The performance of Networks IP and IPN is lower than that of Network I because we apply the PSF and the noise to the training data. The correlation coefficient between the network estimated velocity and the simulation is 0.64, and the R2score is 0.42. In addition, the two-dimensional histogram is shown in Fig. \ref{hist_obs}.
For this network, applying PSF alone does not change the final performance much, although it will change the update speed of the network. This decrease in estimation performance is due to the resolution of observation. Currently, we cannot observe the small-scale flow achieved in the numerical simulation. If the resolution is improved with the development of observational technology, the correlation coefficient between network estimation and simulation will be better.

\section{Comparison with LCT}
We compare the velocity evaluated by the Fourier LCT (FLCT) with the velocity estimated using the network that can apply to the observations made in the previous section. We use the FLCT code by \citet{2008ASPC..383..373F} for LCT.  The FWHM of the Gaussian of the FLCT is 1200 km. 
We can obtain horizontal velocity maps from two consecutive images using LCT. Note that LCT can be performed at some arbitrary interval; we choose 30 seconds as the interval in this study. 
A total of 19 horizontal maps over 10 minutes obtained with LCT are averaged and compared with the result of the network. We have performed 19 time averages under the same conditions for the network evaluation. 
We carry out a parameter survey to optimize free parameters in applying LCT. We test LCT using parameters
that would cover typical temporal and scales of thermal convection with simulated velocity. These parameters included averaging times of 5 min, 10 min, 20 min, 30 min, and snapshot; FWHM values of 300 km, 600 km, 1200 km, and 2500 km; and intervals of 30 s, 60 s, and 120 s. We tested a total of 75 combinations of these parameters. We present a figure on this comparison in Appendix \ref{app-A}.
We show the results of the best parameters that obtained the highest correlation coefficient between the horizontal velocity field from the simulation and the one estimated by LCT.

First, we apply LCT and network to the simulation data and compare their performance.
\begin{figure}
  \centering
  \includegraphics[width=\linewidth, clip]{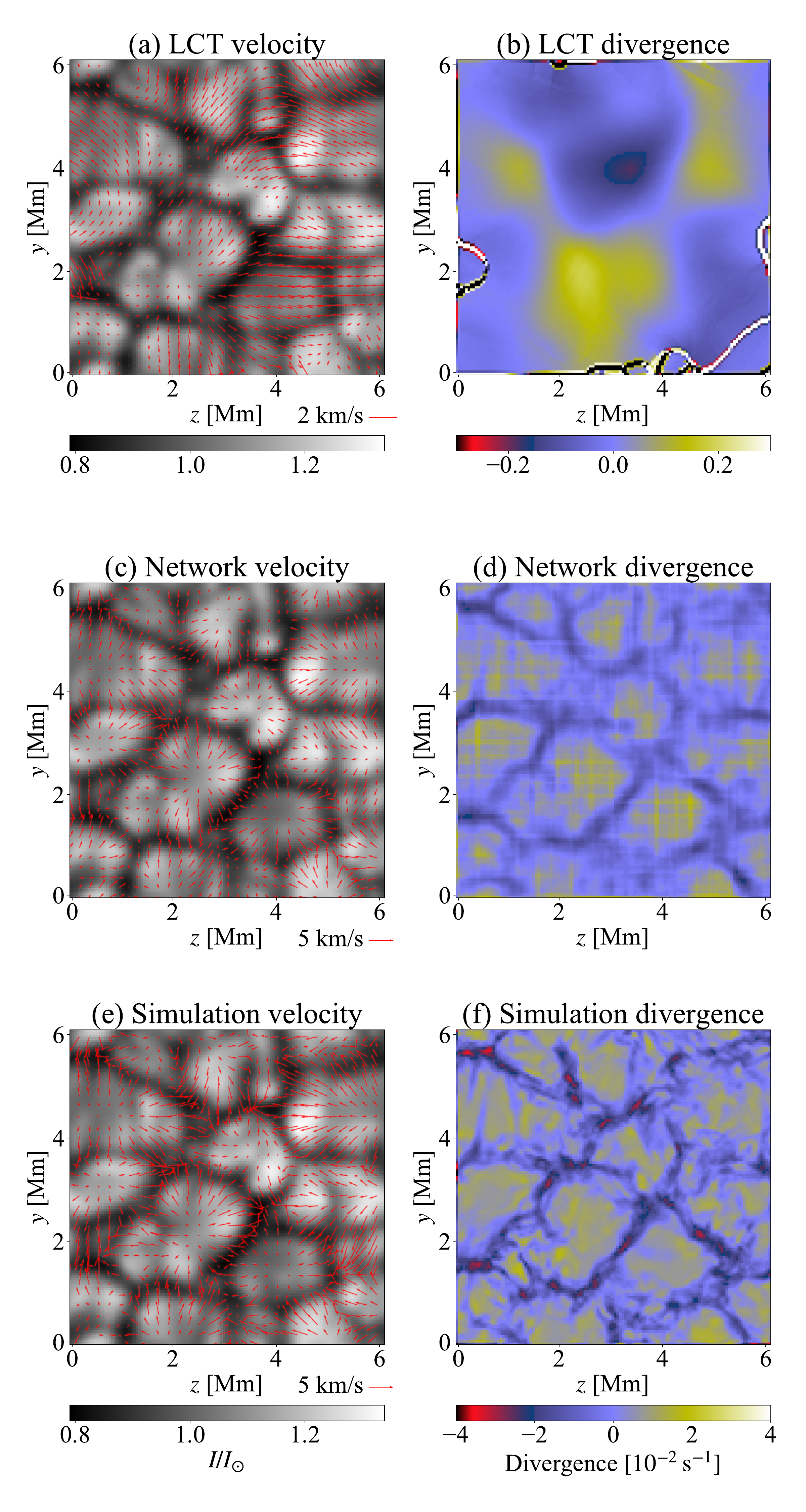}
  \caption{The results of LCT and network for the intensity of the simulation are shown.
  The background in the left column shows the radiation intensity, and the red arrows show the averaged velocity field. The figures on the right show the divergence of the averaged velocity.
  (a) and (b) The velocity estimated with LCT, (c) and (d) the velocity estimated with our network, and (e) and (f) the velocity in the simulation.
  Note that the size of the arrow legend is different for each image.}
  \label{LCT_sim}
\end{figure} 
Fig. \ref{LCT_sim} shows the velocity fields evaluated by LCT and our network and in simulation. The results of the simulation and the network are similar. It appears that some structure is detected by LCT.
However, LCT results do not match the simulation in scale or structure. The correlation coefficient between the simulation and LCT velocities is 0.19 and between LCT velocities and the network is 0.13.
Given our thorough characterization of LCT performance for a reasonable range of input parameters, we conclude that LCT is incapable of accurately recovering granular-scale flows at the spatial resolution studied here.
The correlation coefficient between the network and the simulation velocity is 0.84. The correlation coefficient between the neural network and the simulation increases because the small-scale complex structure disappears due to the temporal average.

\begin{figure}
  \centering
  \includegraphics[width=\linewidth, clip]{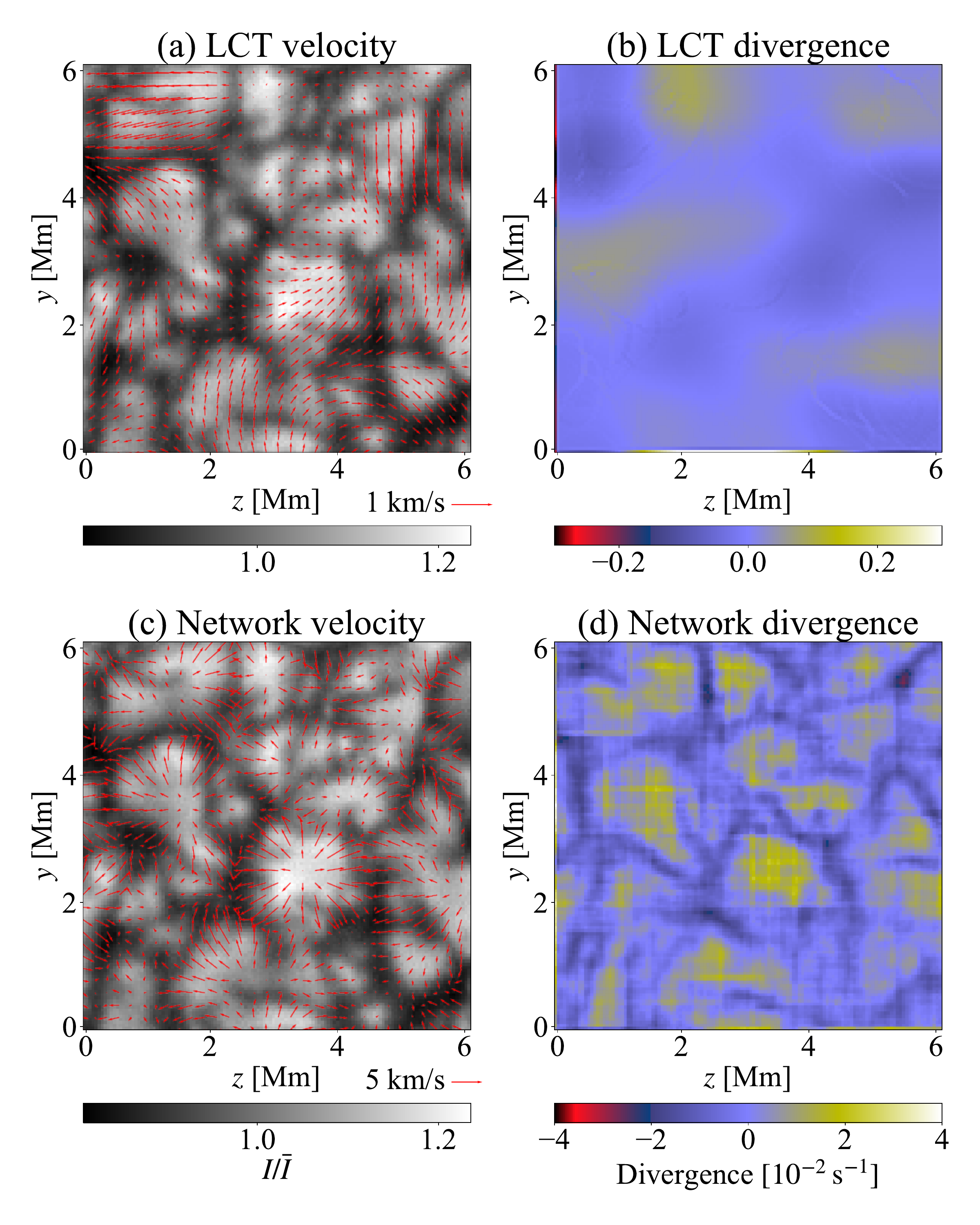}
  \caption{
    The results with LCT and network for the intensity of the observation are shown. The procedure is the same as shown in Fig. \ref{LCT_sim}.
    (a) and (b) The velocity estimated with LCT, (c) and (d) the velocity estimated with our network.
    The radiative intensity is normalized by the mean value $\bar{I}$ over the entire observation area.}
  \label{LCT_obs}
\end{figure}

Next, we compare the results with the observed data. We use the data from 
2007-12-29T11:56:34 to 2007-12-29T12:06:05. 
The results are shown in Fig. \ref{LCT_obs}
. In this case, the correlation coefficient between LCT and the neural network is 0.06, which is hardly consistent.
%
%
With LCT, it is difficult to extract displacements from events with small time and spatial scales.
We cannot capture the flow converging in the granule lane. To effectively capture this flow, resolving this region with a higher pixel count than the LCT window is necessary. 
Observations at higher resolution are therefore required. Considering that pixels in the simulated data (48 km in horizontal extent) approach the highest-resolution observations presently available, higher-resolution instrumentation would have to be developed to employ optical flow methods like LCT to study sub-granular flows.
In addition, because LCT tracks the apparent flow rather than the actual velocity, LCT leads to a smaller velocity than simulation and network velocity. \citet{2018SoPh..293....4M} reported that LCT has difficulty in evaluating the flow in the granulation scale (see also \cite{2018SoPh..293...57T}). 
Tremblay shows it is also difficult for LCT to evaluate the velocity at the edges of the images because occasionally a feature goes out from the data domain in the next step.  Our neural networks do not have this difficulty because only a snapshot is required for the evaluation.

\section{Summary and Conclusion}
We develop a method that estimates the solar horizontal velocity field, which cannot be directly measured, from the radiative intensity and other variables easier to observe using neural network technology. The network is constructed only with convolution. The network can be applied to any image size. Using a GPU, the network can estimate the velocity quickly. Although we cannot obtain the exact horizontal velocity corresponding to the intensity in real observations, we obtained the training and validation data from a numerical calculation using the R2D2 radiation MHD code.

When we include the vertical velocity as additional input, the network performance is improved. By contrast, the vertical magnetic field does not improve the evaluation performance much.

The correlation coefficient between the simulation's velocities and that estimated by the network is 0.83. The overall structure and the velocity inside the granule are consistent. However, it is not easy to estimate the velocity at the granular boundary. High velocities ($>$ 5 $\mathrm{km\ s^{-1}}$) tend to be underestimated. There are still possible ways to improve the evaluation skill of our network. For example, the network can be divided into two networks. When one network evaluates the absolute value and the other evaluates angle, the combination of the two networks may be able to evaluate the complicated turbulent structure well.

Comparing to DeepVel, which uses a similar method to our network, we achieved almost the same performance as the DeepVel with a smaller amount of input data. Our network estimates the horizontal velocity from one snapshot of the image. Thus, we can obtain the velocity from the observation with any length of the time cadence.

If we estimate the horizontal velocity field for the real observation from the intensity using a network trained with simulated data, the network overtrains and does not provide a reasonable result. Therefore, we introduce the PSF of Hinode and reduce the small-scale structure by adding white noise to the training data for the network, which makes the estimated velocities more accurate. Because the network for the observation (Networks IP and IPN) includes the influence from the PSF and the noise, the correlation coefficient is decreased compared with Network I. This implicitly indicates that if the influence of the PSF and the noise is reduced in actual observation in the future, our evaluation ability will increase. 
We apply LCT to the simulation data to determine optimal parameters. We cannot find parameter combinations of similar accuracy to the network estimation.
Because we try to estimate the horizontal velocity in the granulation scale, LCT tends to fail the evaluation. Even for these scales, our network succeeds in evaluating the horizontal flow.
By contrast, evaluating scales smaller than the granulation is difficult because of the loss of information due to the addition of noise. Even training without noise is difficult for small-scale estimation, and \citet{2022A&A...658A.142I} suggest that a significant update is needed, such as changing the training method and increasing the amount of input information.

For future studies, we consider estimating other quantities that are difficult to observe, such as the horizontal magnetic field or the quantity inside the sun, by using the intensity and other observable quantities.

\section*{Acknowledgements}

The results were obtained using Cray XC50 at the Center for Computational Astrophysics, National Astronomical Observatory of Japan.
This work was supported by JST, the establishment of
University fellowships towards the creation of
science technology innovation, Grant Number JPMJFS2107.
This work was supported by MEXT/JSPS KAKENHI (grant no. JP20K14510, JP23H01210 (PI: H. Hotta)), 
P21H04492 (PI: K. Kusano), JP21H01124 (PI: T. Yokoyama), JP21H04497 (H. Miyahara)
and MEXT as Program for Promoting Research on the Supercomputer Fugaku (MEXT as Program for Promoting Research on the Supercomputer Fugaku( JPMXP1020230504 (PI: H. Hotta) and J20HJ00016 (PI: J. Makino))
).
YK is supported by JSPS KAKENHI Grant Number JP18H05234 and JP23H01220 (PI: Y. Katsukawa).
R.T.I. is supported by JSPS KAKENHI Grant Number 23KJ0299 (PI: R.T. Ishikawa).
We express our heartfelt gratitude to the referee for their careful review and valuable comments, which greatly enhanced the quality of this manuscript.

\section*{Data availability}
The simulation data and source code of neural networks underlying this article will be shared on reasonable request to the corresponding author.

\bibliographystyle{plainnat}
\bibliography{Reference}

\appendix
\section{Parameter survey of LCT}
\label{app-A}
This appendix presents the parameter survey of LCT.

\begin{figure*}
  \centering
  \includegraphics[width=0.9\linewidth, clip]{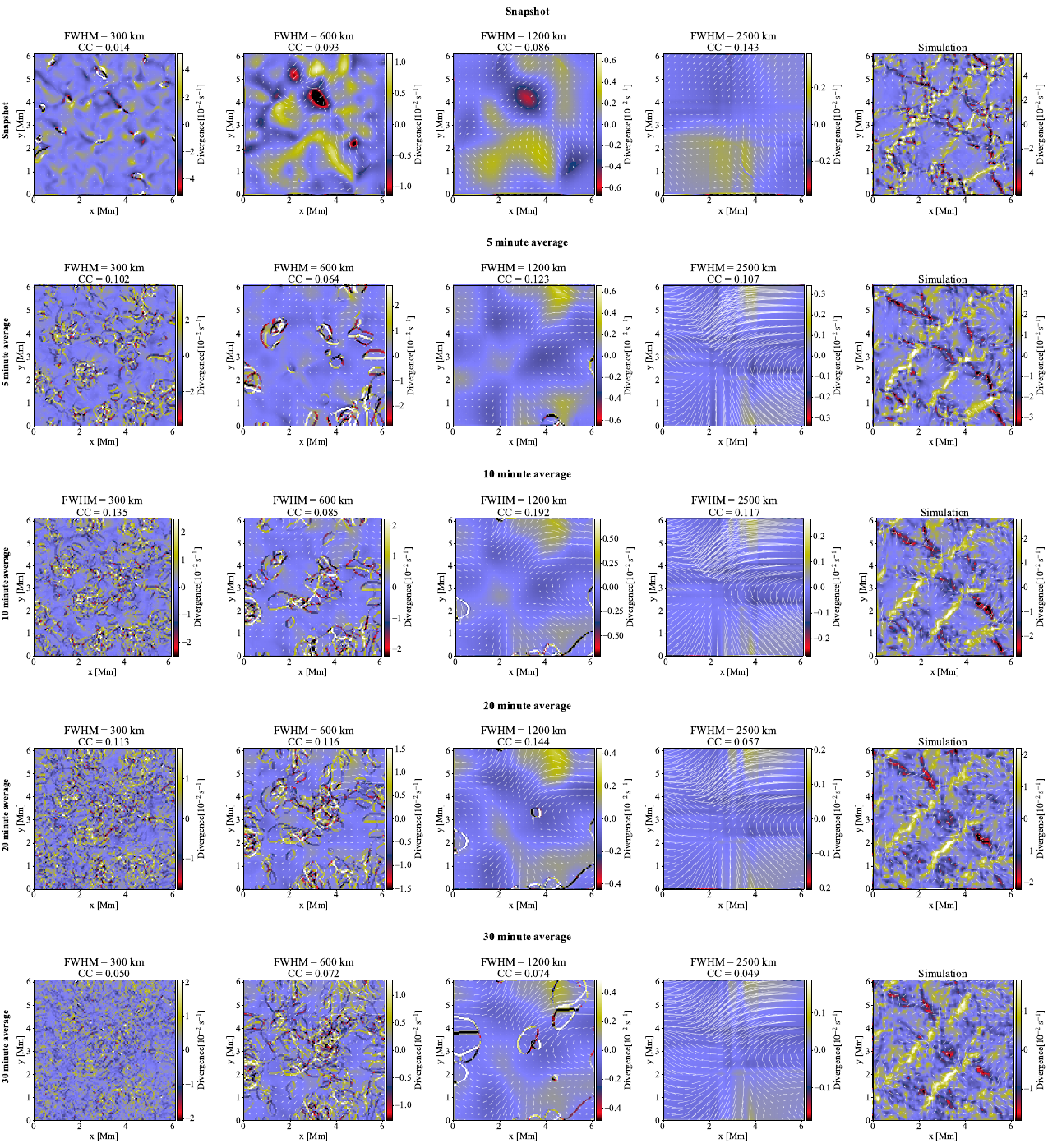}
  \caption{These figures show the divergence of the horizontal velocity field obtained using LCT with a 30 s interval. Each row contains images averaged over the same time; each column represents images with the same FWHM. The rightmost column presents the time-averaged horizontal velocity field of the simulation. Above each image, the correlation coefficient (CC) with the corresponding simulated velocity is shown.
     }
  \label{image30}
\end{figure*}

\begin{figure*}
  \centering
  \includegraphics[width=0.9\linewidth, clip]{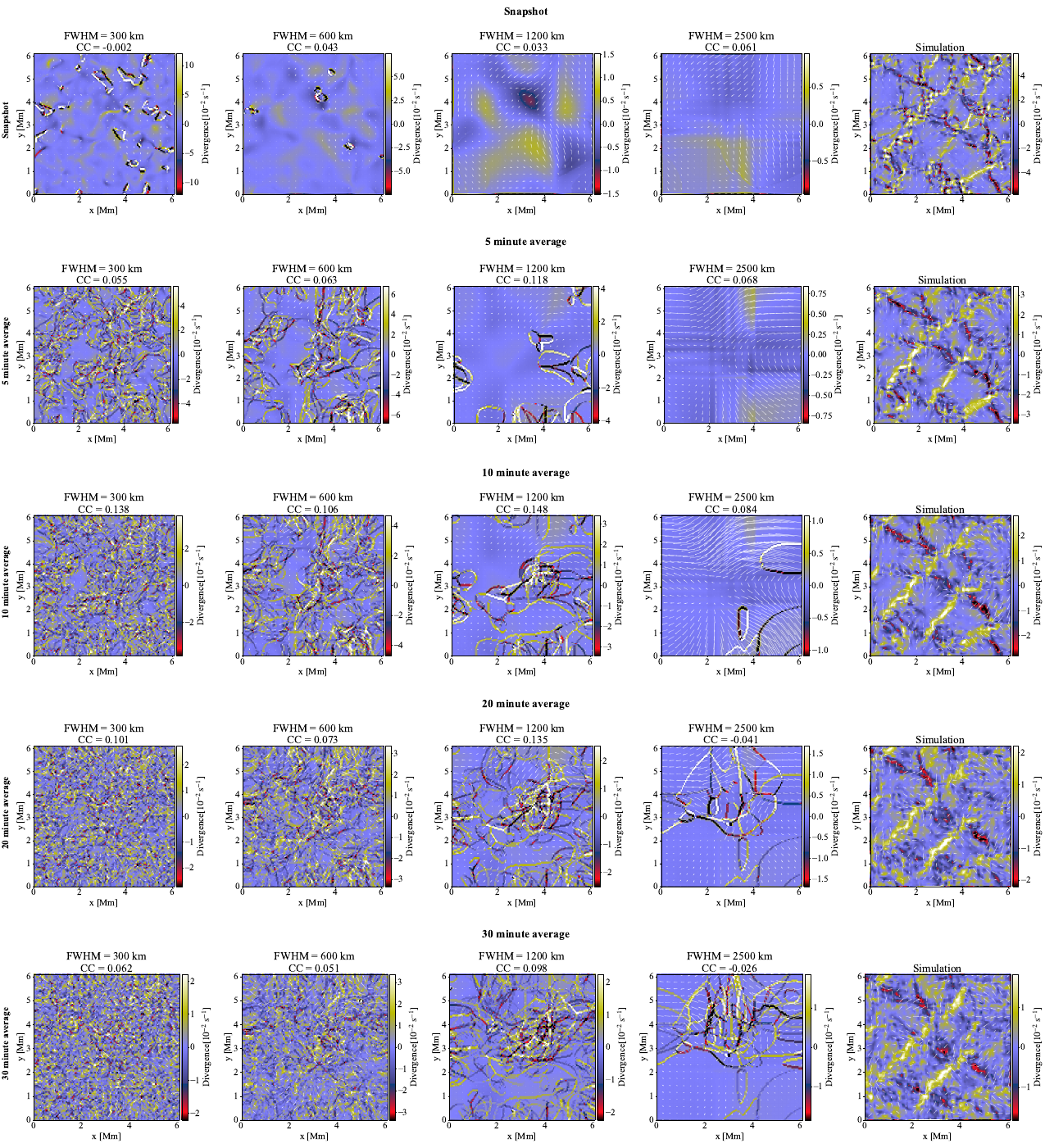}
  \caption{These figures show the divergence of the horizontal velocity field obtained using LCT with a 60 s interval. Each row contains images averaged over the same time; each column represents images with the same FWHM. The rightmost column presents the time-averaged horizontal velocity field of the simulation. Above each image, the correlation coefficient (CC) with the corresponding simulated velocity is shown.
     }
  \label{image60}
\end{figure*}

\begin{figure*}
  \centering
  \includegraphics[width=0.9\linewidth, clip]{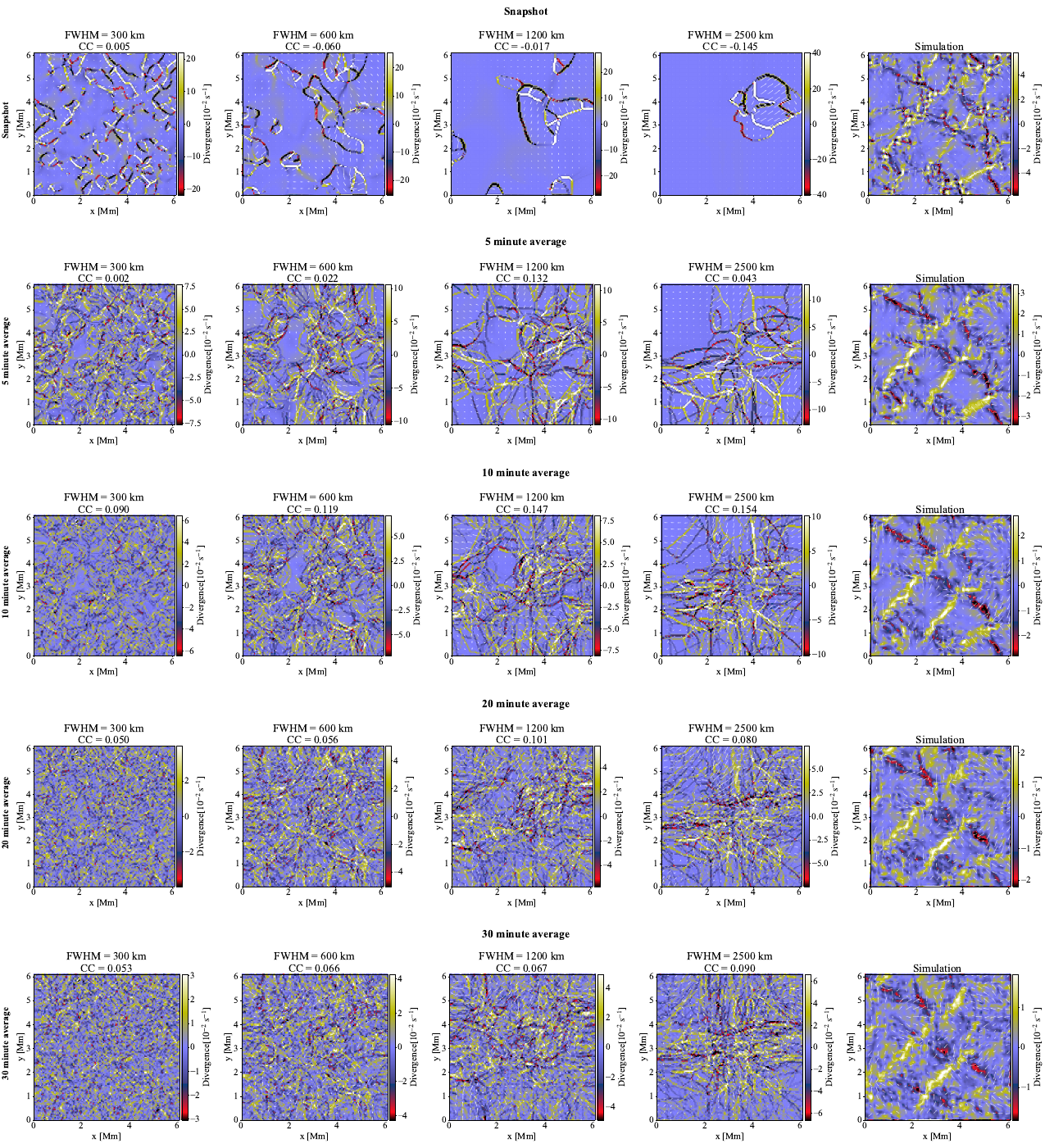}
  \caption{These figures show the divergence of the horizontal velocity field obtained using LCT with a 120 s interval. Each row contains images averaged over the same time; each column represents images with the same FWHM. The rightmost column presents the time-averaged horizontal velocity field of the simulation. Above each image, the correlation coefficient (CC) with the corresponding simulated velocity is shown.
     }
  \label{image120}
\end{figure*}

Figures \ref{image30}, \ref{image60}, and \ref{image120} show horizontal divergence of the velocity field with LCT at 30, 60, and 120-second intervals, respectively. 
The first, second, third, and fourth row show the data with time averages of 5, 10 , 20 and 30 min, respectively. 
The first, second, third and fourth columns show the result with FWHMs of 300, 600, 1200, 2500 km, respectively.
The rightmost column shows the simulated velocity field. The correlation coefficients (CC) calculated with the simulated velocities are shown at the top of each panel. Note that the correlation coefficients are calculated from the velocity field, not the horizontal divergence.

Since CC shows the highest value (0.192) with invervals of 30s, average time of 10 min, and FWHM of 1200 km (third row and third column of Fig. 14), we adopt this parameter in the main manuscript.

\end{document}